# Single crystal growth by the traveling solvent technique: A review

S.M. Koohpayeh

Institute for Quantum Matter and Department of Physics and Astronomy, Johns Hopkins University, Baltimore, MD 21218, USA

Corresponding author: Dr S.M. Koohpayeh, Koohpayeh@jhu.edu
Tel. +1 410 516 7687; Fax +1 410 516 7239

## Abstract

A description is given of the traveling solvent technique, which has been used for the crystal growth of both congruently and incongruently melting materials of many classes of intermetallic, chalcogenide, semiconductor and oxide materials. The use of a solvent, growth at lower temperatures and the zoning process, that are inherent ingredients of the method, can help to grow large, high structural quality, high purity crystals. In order to optimize this process, careful control of the various growth variables is imperative; however, this can be difficult to achieve due to the large number of independent experimental parameters that can be grouped under the broad headings 'growth conditions', 'characteristics of the material being grown', and 'experimental configuration, setup and design'. This review attempts to describe the principles behind the traveling solvent technique and the various experimental variables. Guidelines are detailed to provide the information necessary to allow closer control of the crystal growth process through a systematic approach. Comparison is made between the traveling solvent technique and other crystal growth methods, in particular the more conventional stationary flux method. The use of optical heating is described in detail and successful traveling solvent growth by optical heating is reported for the first time for crystals of $Tl_5Te_3$, $Cd_3As_2$, and $FeSc_2S_4$ (using Te, Cd and FeS fluxes, respectively).

**1. Introduction**

The availability of high quality single crystals is vital for many diverse applications in solid state and quantum electronics as well as for the detailed study of the intrinsic physical properties of materials. To synthesize and grow single crystals of increasingly complex materials with the appropriate properties, crystal quality and homogeneity, growth conditions have to be precisely controlled while growth mechanisms need to be understood at an atomic level. This may be achieved by improving or adapting existing techniques or the development of entirely new methods.

Single crystals have been grown by a variety of techniques from vapor, gaseous phase, melt and solution; however, growth of crystals via solidification from the melt has been the most widely used technique and is generally preferable when possible, e.g. for the crystal growth of the elemental materials and congruently or near congruently melting compounds [1]. Among the bulk growth techniques from the melt, crystallization from high temperature solutions (the flux method) allows single crystal growth of a wide range of materials [1-5]. In this technique, the components of the desired materials are dissolved in a solvent called a flux, and a significant advantage of the method is that the crystals are grown below their melting temperature. This may be essential if the material melts incongruently (i.e. decomposes before melting), or exhibits a phase transition below the melting point (meaning that growth has to be performed at temperatures lower than any such phase transition). Thermal strain can also be minimized in flux growth due to the relatively low growth temperatures and very slow cooling rates.

The flux technique, however, has several disadvantages and one of the main drawbacks is that the growth rate of crystals grown from the flux is generally more than hundred times slower than in melt growth. Furthermore, crystals prepared from the flux using stationary crucibles are typically of small size due to multi nucleation, although the size can be improved using the accelerated crucible rotation techniques [6]. Crystals grown from high temperature solutions frequently develop facets making the linear growth rate difficult to determine as in many cases it is not constant and might be different for each facet, leading to inhomogeneities in the crystal. For industrial production this facetted form can be a disadvantage compared to the round boule one obtains using, for example, the Czochralski or floating zone methods. A further disadvantage is the unavoidable presence of impurities in as-grown crystals in the form of ions of the flux material; this can be either flux substitution for the constituent elements of the desired crystal, or flux inclusions in the crystal. Crystal extraction from the flux can also be quite difficult after the crystal is grown. Methods of removing flux can involve spinning off the molten flux (at temperatures above the melting point of the flux), chemically etching the solid flux away from the crystals, or mechanically removing the crystals from the solidified flux. Regardless of all these disadvantages, the flux technique is popular since the effort required to grow crystals from high temperature solutions may be less than that for melt techniques, while far less in the way of specialized equipment is required.

The ‘traveling’ solvent technique has certain similarities to the flux technique in that crystals of a range of materials can be grown at temperatures below their melting point making it suitable for both peritectic and congruently melting compounds, and crystals of solid solutions of homogeneous composition [7-11]. Following the growth of silicon and germanium crystals from thin solvent zones in the 1950s [7-8], early development of the traveling solvent technique saw the successful growth of GaAs crystals at temperatures much below their congruent melting point using a thin zone of Ga as the solvent [9]. In addition to the advantage of growth at lower temperatures, the crystals grown using this technique are generally larger in size than flux grown crystals via spontaneous nucleation because the traveling solvent technique is in effect a zone melting process in which the existence of a sharp temperature gradient profile along the sample leads to directional solidification. Furthermore, this melt zoning process helps to grow higher purity single crystals since the solvent and other impurities are moved away from the growing crystal [10-12]. Crystals developed from solution (rather than from the pure melt or the vapor) using the travelling solvent are also reported to have more perfect crystalline structures with lower dislocation densities due to a lower solution enthalpy [10].

Notwithstanding the advantages of the traveling solvent technique, optimization of the method is difficult owing to the high number of parameters and their dependence upon each other. Thus, in order to grow higher quality crystals it is vital to study the technique and its relevant experimental variations in detail to attain better control and greater understanding of the growth conditions. The main aim of this article, therefore, is to review the published information pertaining to the various aspects of traveling solvent crystal growth and its associated experimental parameters, while the discussion employs concepts covered in textbooks on bulk crystal growth [1,11], crystal growth theory and techniques [10], and zone refining [12]. In addition, the use of optical heating is discussed in relation to the traveling solvent technique while current experimental practice is illustrated by describing the crystal growth of $Tl_5Te_3$, $Cd_3As_2$, and $FeSc_2S_4$ (using Te, Cd and FeS fluxes, respectively).

## 2. Crystal growth by traveling solvent techniques

Traveling solvent techniques, such as the traveling solvent floating zone (TSFZ), which is container-less, or the traveling heater method (THM), in which a container is used, perform crystal growth from a molten solvent which is typically millimeters long, and has two liquid-solid interfaces. At the melting interface the solvent melts the charge material at a lower temperature than its melting point, the two constituents then being mixed in molten zone; while at the freezing interface a crystal of the same composition as the feed charge is solidified. Since the freezing action rejects the flux into the liquid, the concentrations in the molten zone should not change during the crystal growth process provided that a uniform feeding rate of the charge material is maintained. This differs from the situation in the flux technique where spontaneous nucleation occurs, leading to the growth of relatively small crystals. Rather, in the traveling solvent techniques only part of a charge is melted at a time and the solvent moves away from the growing crystal(s) leading to the formation of larger crystals with higher purity via a directional solidification process. Thermal strain is also minimized in this technique due to the lower growth temperatures and slower cooling rates when compared to other typical melt growth techniques (such as Bridgman, Czochralski or floating zone techniques) that are usually applied for congruently melting compounds.

### 2.1. The traveling solvent floating zone (TSFZ) technique

Among the traveling solvent techniques, the crucible-less technique of traveling solvent floating zone (TSFZ) circumvents the problems resulting from the use of crucibles, in particular, pick up of impurities from the crucible material. It has now become probably the most popular method for the growth of materials with high or very high melting points for which no crucible material is available. In the TSFZ technique, the common methodology is to melt a small 'solvent disk' of different composition between vertical cylindrical feed and seed rods, which have the same composition as the required crystal. Since the molten solvent is supported against gravity essentially by the surface tension of the liquid, the molten zone stability plays an important role in this technique. The molten zone then retains its composition as zoning progresses since the solidification of crystal from the molten zone is balanced by introduction of feed rod material into the zone. Using the TSFZ technique, a single crystal can be generated by spontaneous nucleation or by using a single-crystalline seed crystal as the initial part of the rod, similar to the floating zone (FZ) technique [13].

When compared to the stationary flux technique, crystals grown by the TSFZ technique are typically of higher purity, larger, and they can be extracted easily because there is no crucible or flux around the crystal. However, it should be noted that the TSFZ technique is often more complicated and difficult to control, while the equipment is generally more complex and expensive. Following recent developments, optical heating furnaces are now widely used due to their adaptability, design and high efficiency, thus TSFZ has been utilized to grow a variety of bulk crystals [13]. As is the case for a typical floating zone crystal growth process, the principal growth parameters (feed rod characteristics, growth rate, growth atmosphere and its pressure, temperature gradient, molten zone temperature and rotation rate) can all play key roles during growth of single crystals using image furnaces. The effect of these parameters becomes even more acute in the growth of materials that have 'difficult' intrinsic properties such as high density, low surface tension or a complicated phase diagram. In this regard, more information is given in a review article on experimental procedures using the optical floating zone technique [13]. Summaries of the crystals grown by this the TSFZ technique using optical heating have been reported elsewhere [11,13].

### 2.2. Conventional traveling solvent technique

In contrast to the TSFZ technique, the conventional traveling solvent crystal growths performed using a container do not have some of the difficulties associated with the container-less TSFZ such as molten zone instability, feed rod preparation, molten zone shape, viscosity of the molten solvent, density, etc [13]. The main variant of the traveling solvent growth uses the traveling heater method (THM), whereby a single crystal is grown in a vertically held container by translating a heated sector and thereby slowly moving a molten zone of different composition (solvent) through a polycrystalline charge of the desired crystal. Despite the commonly used term 'traveling heater method', movement of the solvent zone is not necessarily provided by a mechanical movement of a heater, rather the heater is sometimes held stationary and the container is moved in relation to it [10-11]. A less used variant of the traveling solvent techniques is the so-called traveling solvent method (TSM) for which the solvent zone thickness is typically smaller than THM [10]. Given the ambiguities of terminology, in this article the more rational term of 'traveling solvent technique' will be used regardless of the zone thickness or whether translation of the solvent zone comes about by moving the heater or the container. Most of the information provided here can be applied to either variation of the technique.

More information on other less frequently used configurations of the traveling solvent, mostly in a vertical arrangement, such as sublimation, multipass, and cold traveling heater methods, have been discussed in the literature [11].

**2.2.1. Traveling solvent growth using optical heating**

Several heating methods can be employed for the traveling solvent technique, e.g. induction, resistance or laser heating. In particular, the use of light heating (i.e. the focused infrared light radiation from halogen/xenon lamps onto the sample or/and crucible) makes the technique suitable for both conducting and non-conducting materials. This feature distinguishes optical heating from melt zoning using induction heating which can be limited to conducting samples, or samples within a conducting container [13-14]. Fig. 1 shows schematic of zone movement process in the travelling solvent technique using the optical heating.

Another advantage of focused radiation and having a relatively sharp temperature gradient in zone melting is to make short zones feasible with an associated enhanced control of growth fluctuations [10]. Contamination by containers can also be minimized if the container is optically transparent (such as quartz) due to lower heat absorption resulting from optical radiation. However, although light heating is particularly convenient and efficient for many materials that easily absorb light (e.g. oxides and semiconductors), difficulties can arise if the charge substance exhibits a high reflectivity to radiation in the infrared region, as is the case with some metallic samples. A solution to this difficulty may be to contain the charge in a radiation-absorbing container so that the radiation absorber can heat the charge. Since it is difficult to measure the exact temperature using the optical heating system [14], the required operating temperature should be set by adjusting the lamp power. The feasibility of optical heating for traveling solvent experiments using a double ellipsoid mirror furnace is reported for the growth of GaSb crystal with Ga flux in a quartz ampoule [15]. Following that, optical heating was applied to the traveling solvent growth of CdTe [16], $In_{1-x}Ga_xP$ [17], InP and GaSb [18], $Al_xGa_{1-x}Sb$ [19], GaAs [20], and $AgGaS_2$ [21].

One possible advantage of induction heating over light heating is its ability to stir the liquid, the amount of stirring being dependent upon factors such as frequency, coil arrangement and zone size [12]. However, temperature control using induction heating is similarly difficult although this heating mode has proved advantageous for growth of materials such as Si, SiC, GaP, and ZnO [10]. Other complications with induction heating can arise, for instance, if the electrical conductivity of the solvent is higher than that of the growing crystal. This can lead to a higher heat generation and an enhanced convection within the solvent zone which may be accompanied by growth fluctuations leading to impaired crystal perfection. The latter can be alleviated by keeping the molten zone thickness narrow and growing a smaller diameter crystal. Such effects can be lessened when using resistance heating for which the temperature control is more straightforward. Electrical resistance heating has been successfully used in the traveling solvent growth of a great number of materials such as GaAs, ZnTe, and CdTe [10].

### 2.2.2. The choice of growth container or crucible

For the growth container or crucible, it is essential to find a material that is inert to the melt and contaminates the melt as little as possible. Potential contaminants may be agents on the surface of the container, gases from interstices within a porous container, impurities in the container material, or the material of the container itself. The design of a container is also important in order to optimize control of the zone length and zone spacing. For instance, a cylindrically shaped container ensures the consistency of the crystal diameter and uniformity of the in-plane temperature gradient. Additionally, the vertical attitude of the container, where the melt-solid interface is normal to the growth axis, encourages the formation of a more homogeneous crystal because the heat flow, convective flow patterns and solute distribution are all more symmetric [22].

In general, the longitudinal thermal conductance of the container should be less than or comparable to that of the charge. At the same time the lateral conductance, through the container wall, should not be limited unduly. A thin wall container is good on both counts [12]. Among potential crucible materials, quartz glass has advantages such as its incredible thermal shock resistance, low coefficient of thermal expansion, optical transmission properties and good chemical resistance. Sealing a quartz ampoule crucible can also be advantageous since the growth atmosphere, pressure and vaporization can be controlled, while it also provides a protective environment for the samples containing volatile materials. Alternately, a crucible of a more suitable material can itself be sealed within a quartz ampoule.

The use of a vertical container, however, must be approached with caution, as the zone melting and crystallization processes can crack a sealed vertical tube made of a brittle material. Here we can distinguish two possible scenarios; firstly if solid (a pre-melted void-free polycrystalline feed material) ‘contracts’ on melting, a void can form together with the molten zone at the bottom of the tube. Such a void accompanies the molten zone as it moves upwards through the charge and finally bubbles out at the top. However, if the molten zone starts at the top and moves downward, then, solidification begins at the surface of the liquid (below that of original solid due to contraction), and when the molten zone reaches the bottom of the tube, it will expand on freezing and may crack the tube [12].

Conversely, for a material that ‘expands’ on melting the safer operation would be to move the molten zone downwards from top to bottom. This is because the first solid to freeze will, ideally, form at a level slightly above the height of the original solid. Therefore, when the molten zone reaches the bottom of the tube, it contracts on solidification, leaving a void. Although moving the zone from top to bottom helps to give a safer growth, for some materials initiating a single grain might be more difficult to achieve. In this case, in order to perform a growth from bottom to top, use of a porous and non-compacted feed material may be found to accommodate the expansion on melting; however, the potential accumulation of voids within the molten zone can then be detrimental [10].

There are also other factors that can cause cracking. Often, if the solid sticks to (and wets) the wall of the container, a large differential thermal contraction between the growing crystal (solid) and the crucible (container) may cause the container or the crystal to crack [12]. In this respect, it is reported that a greater thermal contraction of the crystal than that of the crucible can help the crystal contract away and detach itself from the crucible wall as both cool to room temperature [22]. Sand blasting of silica ampoules or the formation of a carbon film to obtain a roughened surface has been reported to reduce wetting [22]. Vaporization from the zone and condensation elsewhere can also cause cracking.

### 2.2.3. Growth of incongruently melting compounds

The lower melting temperature of the charge when using the traveling solvent technique significantly facilitates the growth of materials with phase instabilities near the melting temperature. Fig. 2 shows a typical constitutional, or phase, diagram for a binary system in which crystals of the compound x cannot be grown by cooling a stoichiometric mix of A and B phases since x does not melt congruently. This compound, however, can be grown out of excess B. For instance, if a liquid or melt of composition $C_L$ is slowly cooled, just below the liquidus line, it begins to freeze a crystal of solvent of concentration $C_x$ after close equilibrium is approached between regions of the liquid and solid very close to the freezing interface. During solvent traveling crystal growth of compound x, the ratio of $C_x/C_L$ should stay nearly constant (due to effective segregation coefficients being close to unity) in order to grow a chemically homogenous crystal of the same composition as the feed charge. For instance, crystals of $CeSb_2$ and $Ce_2Sb$, which are non-congruently melting compounds in the binary Ce-Sb system, can be grown by the traveling solvent (out of excess Sb and Ce fluxes, respectively) rather than the stationary flux technique [3].

In the $Sc_2S_3$-FeS pseudo-binary system [23], the high melting point compound $FeSc_2S_4$ is an incongruently melting material (at temperatures around ~1822 °C) for which the traveling solvent technique can be utilized. Using an optical heating furnace (Crystal System Corp., Model: FZ-T-4000-H-VII-VPO-PC, having four 1kW halogen lamps) successful crystal growth of $FeSc_2S_4$ out of excess FeS was performed in the author's laboratory by the traveling solvent technique. This was achieved using locally focused optical heating onto a high melting point and inert crucible of pyrolytic boron nitride (PBN) sealed in a transparent quartz ampoule. During the growth, the quartz did not reach excessively high temperatures due to its low light absorption, whereas with electrical resistance heating, its temperature would have exceeded 1200 °C, a temperature at which quartz softens.

For this growth, both the pure polycrystalline charge ($FeSc_2S_4$ powder, ~300 mg) and solvent (FeS powder, ~150 mg) were placed in a PBN crucible with the flux having a lower melting temperature at the bottom of the container while no seed crystal was used. (The use of such small amounts of feed material was due to the long preparation process for high purity $FeSc_2S_4$ powder.) The inner diameter of crucible was 10 mm, the wall thickness was 1 mm, and the bottom of the container where the growth was initiated was shaped as a cone (over a length of 20 mm) to provide the conditions for competitive growth and formation of a single grain. The crucible then was sealed in a quartz tube (under 0.3 bar argon), vertically placed in the furnace, and monitored using a camera fitted in outside the growth chamber. The focused radiation at 80% lamp power melted the solvent which was then moved upwards along the polycrystalline $FeSc_2S_4$ powder by the translation of the optical heating source at a rate of 0.3 mm/h. Rotation rates of 5 rpm were employed for the crucible, and only one zone pass was performed.

Despite limited availability of the pure polycrystalline charge material, the first reported high quality and stoichiometric $FeSc_2S_4$ single crystal was successfully grown, as shown in Fig 3. Powder X-ray diffraction (XRD), using a Bruker D8 Focus X-ray diffractometer operating with Cu $K\alpha$ ($K_{\alpha 1}$ = 1.5406 Å) radiation, confirmed the cubic structure (space group $Fd\bar{3}m$) for the crystal, and the room temperature lattice parameter of $a$ = 10.51288(3) Å was measured using the Bruker TOPAS software (Bruker AXS).

### 2.2.4. Growth of congruently melting compounds

The traveling solvent technique can also be applied to congruently melting materials for which crystal growth using conventional melt techniques at temperatures near their melting points could lead to problems such as excessive oxidation, high vaporization rates or high temperature phase transitions. For instance crystals of the congruent, high melting point compound CeSb ($T_m \approx$ 1750 °C) [3] can be grown out of either Sb or Ce flux as one of its constituents at temperatures around 1500 °C using the traveling solvent technique. High quality and pure CeSb crystals have proved difficult to grow via the flux method using Sn as the flux, due to formation of second phase inclusions, and container problems when using quartz above 1200 °C [3].

To give another example, the vapor pressure of Arsenic (As) rises with temperature from 1 bar at ~610 °C to above ~37 bar at 818 °C (near the melting point) [12,24]. Therefore, although an excess of the volatile As can be added in an attempt to provide the ambient vapor pressure desired during a high temperature growth of a compound with As content [12], growth at a lower temperature is clearly advantageous. In this context, crystal growth of the compound $Cd_3As_2$ was attempted in our laboratory using the same optical heating furnace as utilized for the growth of $FeSc_2S_4$. Based on the binary phase diagram for Cd-As [25], $Cd_3As_2$ is reported to be a congruently melting compound that shows different phase transitions upon cooling to room temperature. Despite the congruently melting nature of the high temperature phase $Cd_3As_2$ at ~721 °C, the growth of the room temperature phase at temperatures less than 500 °C should only be possible out of excess Cd solvent in order to circumvent both the phase transitions and excessive As vaporization at the higher temperatures.

For this purpose, 1.35 g of polycrystalline $Cd_3As_2$ was placed on top of 0.71 g of Cd flux in a quartz tube (of 6 mm inner diameter and 1 mm wall thickness, which was shaped as a cone at the bottom part) and sealed under ~0.25 bar argon. To avoid possible cracking and toxicity issues, this quartz tube was re-sealed in a larger quartz tube (of 10 mm inner diameter and 1 mm wall thickness), and then was held vertically in the optical floating zone furnace. Using about 23% of the lamp power the solvent was melted, while the use of a transparent quartz container and a directly focused radiation/heating onto the sample meant that the molten zone and growth process could be observed by a camera. The molten solvent was then moved upwards at a rate of 1 mm/h and the quartz container was rotated at the rate of 10 rpm leading to the growth of a high quality and stoichiometric crystal of $Cd_3As_2$ (Fig. 4a). Thus the use of the traveling solvent technique and application of Cd as a solvent eliminated the need of any high pressure As atmospheres during the crystal growth process. Moreover, the use of this technique significantly helped to control vaporization and avoid high temperature phase transitions [25-26] by lowering the melting point from ~721 °C to less than ~480 °C. The phase purity and the crystal structure (Space group $I4_1/acd$ [27]) of the crystal was confirmed by the powder X-ray diffraction (Fig. 4b), and the lattice parameter were measured to be a = 12.6515(2) Å and c = 25.4511(8) Å using the Bruker TOPAS software.

Another example of a congruently melting compound for which the traveling solvent technique using optical heating yielded satisfactory crystal growth is $Tl_5Te_3$. Although the phase $Tl_5Te_3$ is reported to be structurally stable over a range of compositions (from 62.5 to ~ 65.0 at% Tl) [28], the physical properties and lattice parameters are very sensitive to small changes in composition [29-30]. Therefore, precise control of the stoichiometry of the grown crystal is essential, as only the stoichiometric crystal (with the ratio of 62.5 at% Tl and 37.5 at% Te) superconducts ideally. In this regard, direct solidification from the molten phase of $Tl_5Te_3$ does not necessarily produce the stoichiometric crystal, depending on the growth conditions. Moreover Te vaporization at temperatures around the melting point of $Tl_5Te_3$ (~443 °C) can also deteriorate the growth process. However, crystal growth of the superconducting phase of the $Tl_5Te_3$ compound was successfully performed using 2.5% excess Te as a flux in a quartz container, similar to the experimental setup used for the $Cd_3As_2$. Fig. 5 shows a picture of the resulting high quality and large crystal of $Tl_5Te_3$, together with a powder X-ray diffraction confirming the phase purity of the crystal. Lattice parameters of a = 8.9245(2) Å and c = 12.6102(4) Å were measured for the grown crystal, consistent with powder diffraction of stoichiometric polycrystalline $Tl_5Te_3$.

**3. Other influential growth parameters during the traveling solvent process**

The conditions of solidification and growth (e.g. thermal gradients, interface velocities, degree of turbulence in the liquid, etc.) are of great importance since they can change the equilibrium needed to maintain a uniform compositional difference ahead of the growing crystal. Although analysis and modeling of mass and heat transport can be of great help in determination of the growth conditions [11], this subject, in general, is very complex due to its essentially transient nature, and seems to be one of the major factors limiting the yield of bulk crystals grown from the melt by the traveling solvent technique. A brief review of the most important parameters is given in this section.

**3.1. Seed crystal**

One of the factors that can limit the use of this technique is the availability of high quality seed single crystals which are the preferred method of preventing multi nucleation from the initial stage of the growth. In this process, the seed crystal is partially melted to provide a new growth interface from which the crystal growth is started. Such seeding is particularly important in commercial production of larger size crystals, while seeding also helps to grow crystals of a particular crystallographic orientation. Ideally the seed crystal, which is usually placed under the solvent depending on the growth configuration, should be of the same composition as the crystal to be grown, although in some cases it is possible for it to be different [11,31]. A seed of smaller diameter than the required crystal can be used in the traveling solvent technique by employing tapered crucible ampoules [32].

The seed can be obtained by other techniques or by performing a preliminary unseeded traveling solvent run, although in some cases, the use of a crucible with a pointed bottom might eliminate the necessity for a seed crystal [10]. However, the advantage of seeding in the traveling solvent technique has been reported for the crystal growth of various materials. For instance, the reproducible growth of oriented

PbTe crystals, free of low angle grain boundaries, is reported using a single crystal seed [33], while crystal growth improvement of $CuGaSe_2$ using the seeding technique is also reported [34].

The seeding position in the furnace when using opaque crucibles can be difficult to ascertain since the seed cannot be observed to perform a precise melting process; therefore, a prior detailed thermal analysis of the system is needed. When using the optical heating technique with a transparent container (e.g. quartz), however, seeding can be easily performed, as reported in this article.

**3.2. Traveling rate of the melt-growing crystal interface**

During the traveling solvent process the solute (of the same composition as the material to be grown) is preferentially solidified at the advancing liquid-growing crystal interface, forming a boundary layer in the vicinity of the growth interface over which all the transport and significant concentration variations take place by diffusion [12]. The nature and thickness of this layer, which is very often difficult to determine, can change by variations of the growth conditions, such as the degree of stirring and mixing, the diffusivity of the material, the viscosity of the liquid, the conditions of fluid flow, and the travel rate of the solid-liquid interface. Among these factors, the growth rate (i.e. the traveling rate of the melt-growing crystal interface) is a critical parameter since the compositional differences necessitate solute diffusion at the solid/ liquid interface, which generally takes place slowly and therefore limits growth to a correspondingly slow rate. For instance, the diffusion constant in the liquid zone in the case of GaP was found to be of the order of $10^{-4}$ $cm^2$/sec, meaning that the growth rate could not exceed the value of about 5 mm per day [10].

While the growth rate values for a variety of materials grown by the traveling solvent technique are mostly within the range 1 to 7 mm/day [10-11], an extremely low growth rate of less than 0.01 mm/day has been reported for $CaWO_4$. When the rate was increased to 0.07 mm/day by increasing the temperature gradient, this resulted in break-up of the solution zone due to the constitutional supercooling inherent in the traveling solvent technique [10]. Solvent trapping into the growing crystal can also occur if the movement rate of the heating source exceeds the maximum growth rate. In some cases, higher growth rates might be possible when a combination of favorable growth conditions (e.g. higher temperature gradients, convection, etc.) is applied [10]. Lower growth rates, in general, are necessary to avoid constitutional supercooling and interface breakdown. Constitutional supercooling leads to instability near the liquid-growing solid interface which can cause the incorporation of impurity inclusions into the growing crystal, mosaicity, polycrystalline growth, or even termination of growth [10].

**3.3. Temperature variations, transport phenomena and compositional homogeneity**

Control and optimization of the temperature and the magnitude of its gradient across the solvent zone is of great importance in order to perform a successful traveling solvent crystal growth. Variations of both axial and radial temperature distributions affect the experimental conditions during an actual growth [10-11]; for instance, closer control of melt temperature and thermal symmetry can help to eliminate unwanted variation in molten zone composition. This can be achieved by rotating the crucible to average out temperature asymmetries [22]. In general, the sample dimensions, thermal conductivity of the material, solvent and crucible, mode of heating, fluid flow and convection within the molten solvent are all important factors affecting the temperature distribution. In a related study, the influence of radial and axial temperature distribution on the formation of Te inclusions in CdTe crystals grown using tellurium as the solvent is reported [35]. Studies of radial and axial temperature distribution through computational modeling and simulation of the heat and mass transport can be very informative, as reported for materials such as CdTe, PbTe and $Ga_xIn_{1-x}Sb$ grown by the traveling solvent technique [10-11].

It is always the case, of course, that the temperature profile along the length of a sample during zoning will never stay constant even if the power setting is unaltered because heat losses at the ends of the sample will differ from those in the central portion. This will be the case in any sample that is not infinitely long and will occur whether or not a crucible is used.

Along with considerations concerning the temperature profile and its symmetric design, the molten zone temperature should be sufficiently high to ensure adequate solubility of the feed material into the solvent to achieve a reasonable growth rate. For instance, during the traveling solvent growth of $Al_{0.1}Ga_{0.9}As$ from an Al-Ga solvent, a molten zone temperature in excess of 1000 °C was found necessary in order to ensure an AlAs solubility sufficiently high to allow for acceptably rapid AlAs diffusion. Similar results were found for the homogeneity of (Ga,In)P crystals grown by the same technique [10]. An increase in temperature gradient above a critical value is also important to prevent compositional perturbations due to constitutional supercooling ahead of the crystal-solvent interface [10-11].

In conjunction with the selection of a suitable solvent material, precise temperature management is vital to the control of stoichiometry and prevention of any detrimental doping effects. In particular, where a specific degree of electronic doping from the solvent is required, it is reported that a decrease in the growth temperature can increase the effects of doping due to either impurities or intrinsic deviations from stoichiometry [10]. The latter, however, may be affected significantly for systems with retrograde solid solubility [10]. In general, for any particular system (that is the crystal material, the selected solvent, doping type, etc.) an optimized growth temperature should be applied in order to grow more stoichiometric crystals with the appropriate doping levels.

Clearly the growth temperature needs to be determined along with the selection of an appropriate solvent. A suitably chosen solvent should have negligible solubility in the growing crystal whereas the polycrystalline feed material (the solute) should have a sufficiently high solubility in the liquid solvent at the growth temperature. It should also have a low vapor pressure at growth temperatures, and good wetting properties with the material to be grown. Meanwhile, the solvent should not form a compound with either the growing crystal or its constituent elements nor react with the container material at the crystal growth temperature.

Rotation of the crucible or growth ampoule also seems necessary to control radial uniformity and to have a steady flow by introduction of a swirl component in the fluid motion which arises from centrifugal and Coriolis forces. The effect of the increased convective transport by a centrifugal force may also be used to increase a growth rate that would normally be low (perhaps a few millimeters per day) due to the limited transport rate of the species in the solution and the formation of solution precipitation [36]. Enhancement of convection at an acceleration of 20 times gravity is reported to have increased the growth rate by about one order of magnitude for the inclusion-free GaSb crystals grown by the traveling heater method [37].

Along with any forced convection, natural convection in the molten zone can also be an effective factor since it mixes the liquid by virtue of localized differences in density. These in turn may arise from differences in concentration [38], or from differences in temperature [39-40]. Natural convection, in general, can cause fluctuation in both the temperature and the melt content along the direction of growth, and as a result can lower the crystal quality and homogeneity. In the long established zone melting and Bridgman bulk crystal growth methods, temperature fluctuation due to convection is reported to be the major source of the non-uniformities [41]. Such inhomogeneities, either on a microscale (e.g. doping striations) or a macroscale (longitudinal and lateral segregations), can be deleterious to the properties of crystals.

Overall, transport phenomena (e.g. fluid motion, heat transfer, and mass transfer) and the interaction of different convection mechanisms in melt crystal growth from solution are undoubtedly complex [36, 41-43]. Therefore, it is important to optimize the rotation schemes for each crucible diameter, melt size, fluid viscosity, heating type and other relevant experimental factors. Additionally, a good design of the growth furnace, with limited thermal asymmetry and pulling device instabilities, is a valuable asset.

### 3.4. The liquid-growing crystal interface

During growth using the traveling solvent technique, the curvature of the liquid-growing crystal interface can be a parameter of considerable importance in determining crystal perfection. Curvature can be affected by the complex heat transfer between various components in the growth system and may vary with crucible rotation rates, thermal conductivity of crucible, temperature gradients in both the melt and growing crystal, encapsulant, gas, sample dimensions, etc. Although such dynamic growth conditions make it hard to directly observe and study the shape and nature of the melt-crystal interface, models of interfaces, their predictions, and the effects on growth and crystal quality have been investigated in several reports [11,13,22,44-46].

Studies of the heat transfer and its effects on the interface shape have been summarized in a recent publication [45]. It has been argued that a convex shape (with respect to the growing crystal) is formed when the heat radially flow into the crucible, while a concave interface is due to the flow of heat to outside. Formation of a concave interface is also supported due to the release of latent heat along the solidification interface toward the cooler exterior. In addition, a concave shape is formed when the thermal conductivity of liquid is larger than the growing crystal, leading to an outward heat flow from the greater temperature of the charge [45]. A larger thermal conductivity along the crucible wall than that of the solid also tends to make the interface shape more concave [22].

In general, planar or slightly convex interface shapes are reported to be desirable [10,22,44,46]. A planar or flat liquid-solid interface makes it less complicated to theoretically develop and study the axial segregation mechanisms, material transport, and compositional profile (e.g. in ternary crystals). A flat interface also helps to obtain crystals of more homogenous composition by decreasing radial segregation associated with the interface deflection [47]. Crucible wall interaction with a convex interface shape, in particular, is reported to reduce the propagation of multigrain nucleation and twins into the growing crystal by forcing such defects outward toward the crucible wall. Such defects are more likely to be extended toward the growing crystal with a concave interface shape [45]. Several experimental studies for the growth of CdZnTe compound also support the reduction of new grain growths due to a convex interface shape [45-46].

When employing optical heating for traveling solvent floating zone (TSFZ) crystal growth, the temperature gradient, opacity of the sample, growth speed, and forced convection in the molten zone are reported to control the shape and nature of the liquid/solid interface [13]. In this technique, the melt-growing crystal interface meets a surrounding gas (rather than a crucible wall), meaning that other factors such as gas type and pressure can play key roles in the heat transfer mechanism [14]. As reported, a sharper temperature gradient can lead to a concave solid/liquid interface, while a reduction in the temperature gradient can give a flat or convex interface [13,44]. However, a very low gradient should be avoided since it can enhance the possibility of constitutional supercooling [44]. The opacity of the sample material can also affect the interface shape, with concave interfaces being more likely in transparent, or nearly transparent, materials, while convex interfaces appear to dominate in opaque materials. Using a lower growth speed seems more likely to give a planar growth front, while a higher rotation rate (forced convection in the molten zone) can be a way of lowering the convexity of the interface, so giving a more stable molten zone.

During TSFZ or the traveling solvent technique crystal growth using a transparent container (e.g. quartz), a camera fitted outside the growth chamber might enable observation of the liquid-solid interface unless the charge itself is opaque (such as metallic samples). For a real time, in depth powerful detection and visualization of a solid-liquid interface, a solid-penetrating radiation source is needed [48]. Depending on the material to be grown, X-rays or gamma rays can be used, and differences in the intensity of the detected energy indicate the shape of the liquid-solid interface. Real time study of convection in the melt, interface morphology and kinetics will provide significant information for optimization and development of both the experiments and analytical/numerical modeling.

**3.5. Volatile constituents and removal of impurities**

The preparation of crystals of compounds having one or more volatile constituents is a complicating factor in many crystal growth techniques and can lead to deviations from stoichiometry. However, since crystal growth using the traveling solvent technique can be performed in a completely sealed tube, vaporization can be regulated by controlling the vapor pressure of a volatile component based on the temperature (of the coolest part of the tube) and an excess of the volatile constituent (added either to the melt or placed elsewhere in the chamber to provide the ambient vapor pressure desired) [22,12]. The rate of evaporation can also be restricted by either introducing an inert gas (e.g. argon) into the growth chamber (tube), or regulating the zoning (or growth) rate of the crystal [12].

Variations in partial pressure during a normal growth process by moving a molten zone from the bottom upwards in a cylindrical crucible are mostly reflected in deviations in the melt composition and ultimately in the crystal stoichiometry and quality, as reported for GaAs [22]. However, in some cases where the interaction of vapor with the growing crystal is necessary, the horizontal traveling solvent method has the advantage of a free surface which can lead to stable growth conditions, as reported for the growth of inclusion-free single crystals of GaSb and InSb [49].

The zone melting process [12], solution zone passages [10-11,50-51] and the choice of an appropriate solvent (e.g. removal of Cu impurity by Te solvent in the growth of CdTe) [10-11] can sweep undesirable impurities away from the growing material, while vaporization can also be used to remove volatile impurities which usually condense on the cold surface of the container [12-13,52]. There are times when an impurity that is not normally volatile can be made volatile through a chemical reaction; for example, by slightly changing the tube atmosphere (e.g. a small amount of oxygen can react with boron, when this is an impurity, to form a volatile oxide which rapidly evaporates before oxidizing the crystal [12].) Depending on the vapor pressure of the material being grown, a clean vacuum environment might also offer the possibility of continuous removal of volatile impurities from the system. Moreover, to prevent re-entry of a volatile impurity into the zone, the use of an open tube system with a flowing inert gas can remove the impurity as vapor. In general, however, a limit on the amount of an impurity that can be removed by vaporization is set by the relative volatility of the host substance. The effect of reaction equilibria and kinetics on the removal of an impurity by evaporation from the melt is reported in more detail by Kraus and Winkler [53].

**4. Summary**

The traveling solvent technique can be applied to grow single crystals of many intermetallic materials, oxides, and other systems of interests that can be difficult to grow by other means due to high temperature phase instabilities (e.g. phase transformation, decomposition near the melting temperature or incongruent melting). The main benefit of the technique is that materials can be grown at temperatures below their melting points, where vapor pressures and decomposition rates are lower. Crystals grown using the travelling solvent technique are potentially larger and of higher purity than crystals prepared, for example, by flux growth, due to the effects of directional solidification and solvent zone passing, respectively. Growth at a lower temperature also decreases the amount of possible contamination from the crucible (if one is used), potentially reduces the non-stoichiometric defect concentration, and gives crystals in which dislocation densities can be orders of magnitude lower than those prepared by other methods.

While the traveling solvent technique has inherent advantages for the crystal growth of certain materials, optimization of the growth conditions and experimental set up is crucial for the successful growth of large, homogenous, pure and high structural quality crystals. Among the factors to be considered are selection of solvent material, the growth temperature and gradient, which determines the size of the molten zone, the zoning (or growth) rate, which determines the feed rate of the charge material into the molten zone, the type of growth atmosphere and its pressure, the shape of the liquid-growing crystal interface and the rotation rate or degree of mixing in the liquid. The choice of these growth conditions will depend upon the properties of the material in question such as its density, surface tension, viscosity, temperature-composition phase diagram, vapor pressure, thermal conductivity, compositional homogeneity and the purity of the feed material. In some cases, factors such as crystal structure, crystal orientation and faceting might also need to be taken into consideration. Furthermore, the experimental setup (e.g. furnace design, heating type, choice of vertical or horizontal configuration of the growth process) also plays a key role during crystal growth using the traveling solvent technique. By following the guidelines discussed above, high quality crystals of $Tl_5Te_3$, $Cd_3As_2$, $FeSc_2S_4$ (using Te, Cd and FeS fluxes, respectively) have been grown using the traveling solvent technique employing optical heating of a vertically held sample.

**Acknowledgments**

The author would like to gratefully thank Dr. D. Fort for his time and helpful discussions. The author also thanks K. Arpino and J. Morey for their assistance. This work was supported by U.S. Department of Energy (DOE), Office of Basic Energy Sciences, Division of Materials Sciences and Engineering under award DE-FG02-08ER46544.

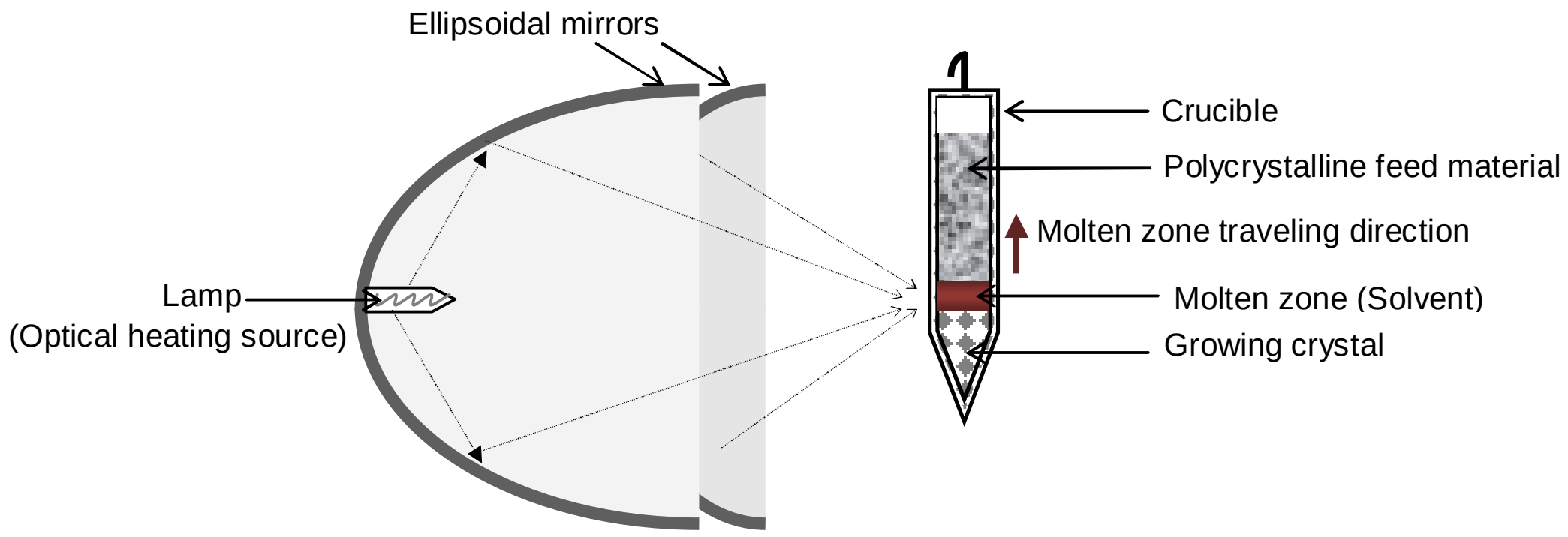

**Fig. 1.** Schematic of zone movement process in the traveling solvent technique using the optical heating. In this technique, the light from halogen or xenon lamp(s) is focused by ellipsoidal mirror(s) onto a vertically held crucible to produce a molten zone, which is then moved upwards along the polycrystalline feed material by translating either the mirror(s) upwards or the crucible downwards.

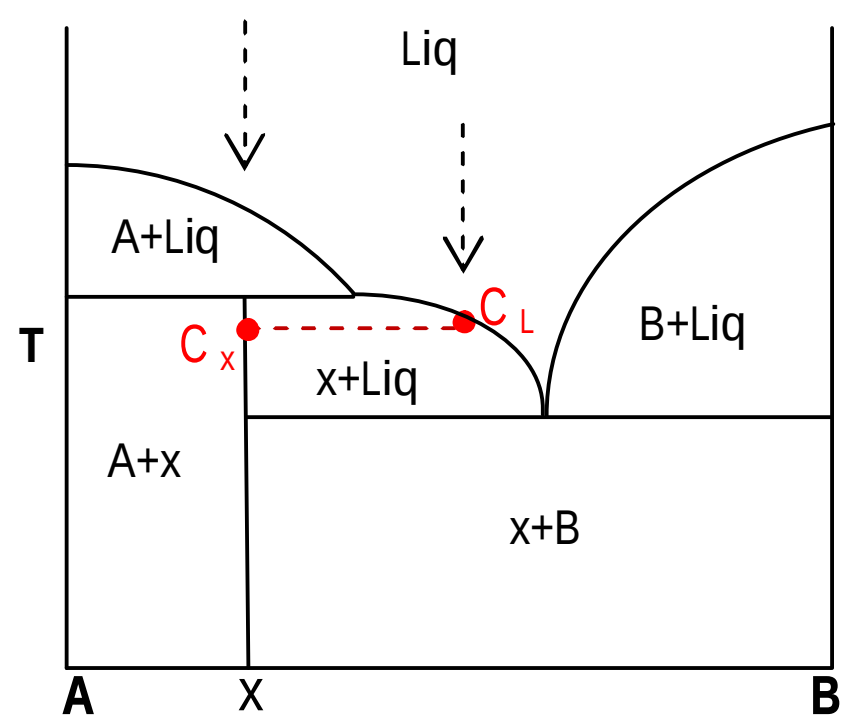

**Fig. 2.** Composition-Temperature phase diagram for a binary system

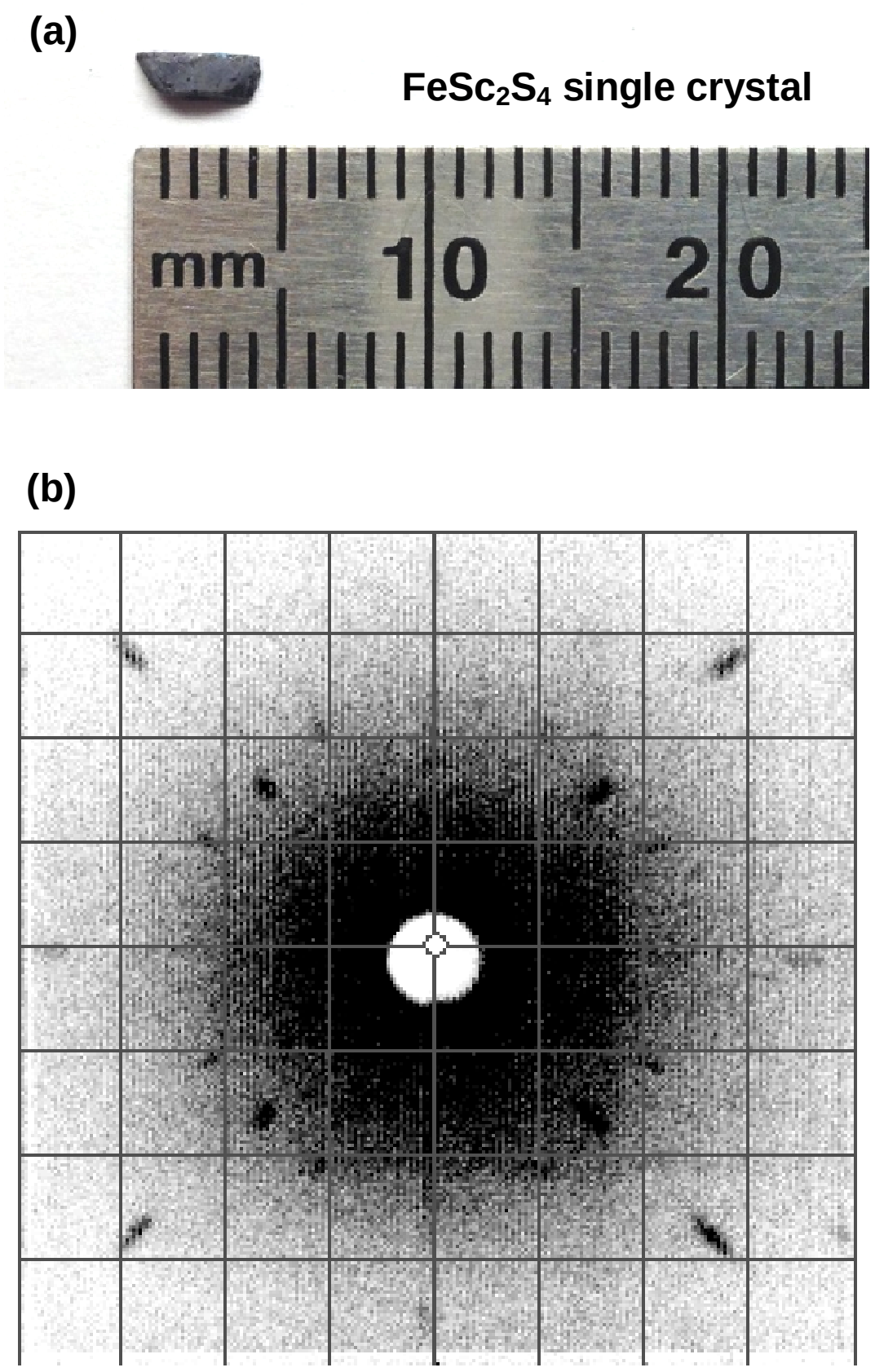

**Fig. 3.** (a) $FeSc_2S_4$ single crystal grown by the optical heating traveling solvent technique using FeS as the flux, (b) ‘a’ plane X-ray Laue pattern taken from the as-grown crystal

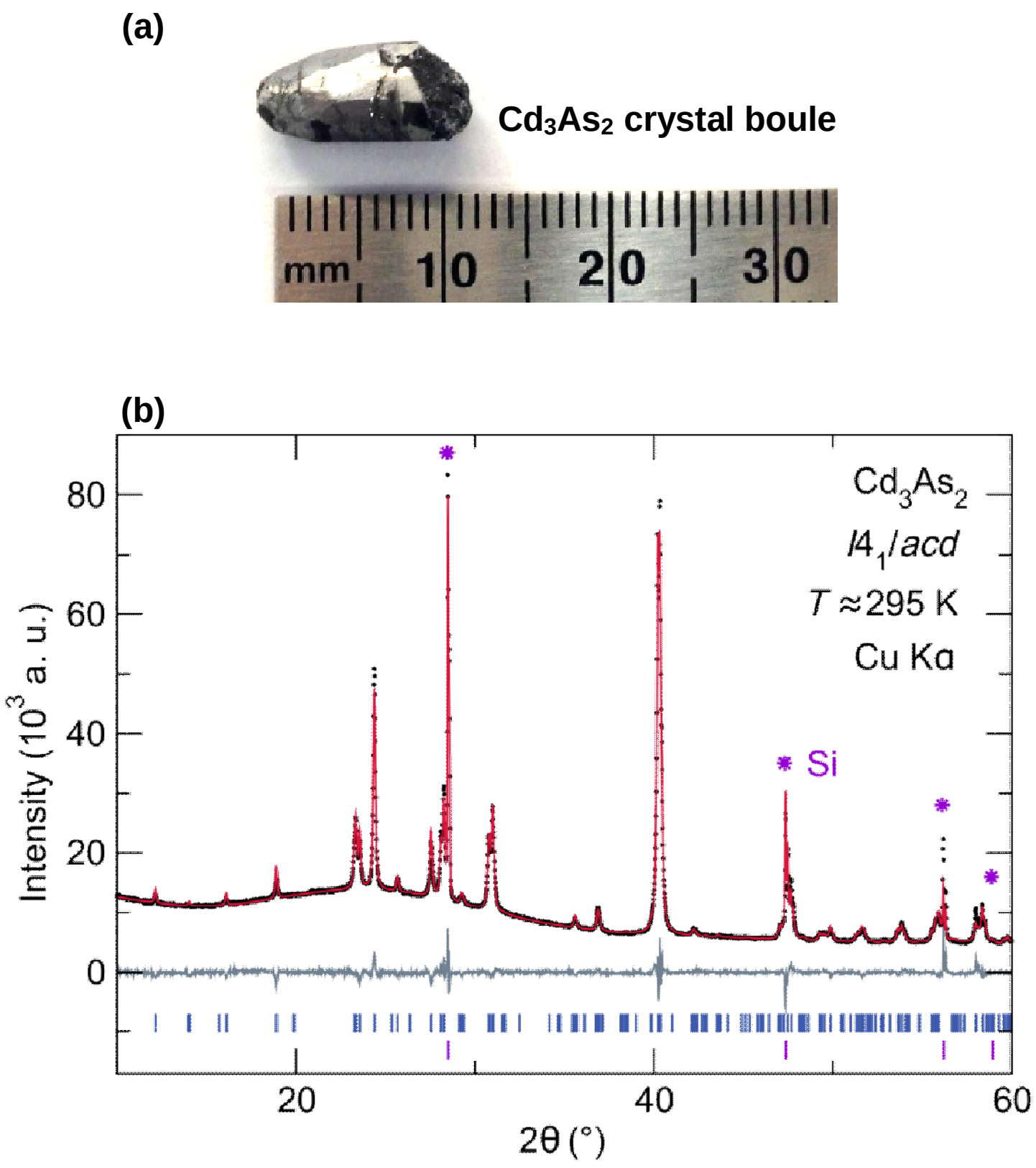

**Fig. 4.** (a) $Cd_3As_2$ single crystal grown by the traveling solvent technique using an optical heating furnace. (b) Powder x-ray diffraction data (black circles) is fit (red line) by Rietveld refinement, with the difference curve in gray. Peaks from a Si standard are marked with purple asterisks. Blue tick marks denote reflections allowed by the $I4_1/acd$ space group of $Cd_3As_2$, and purple tick marks denote reflections allowed for the Si standard.

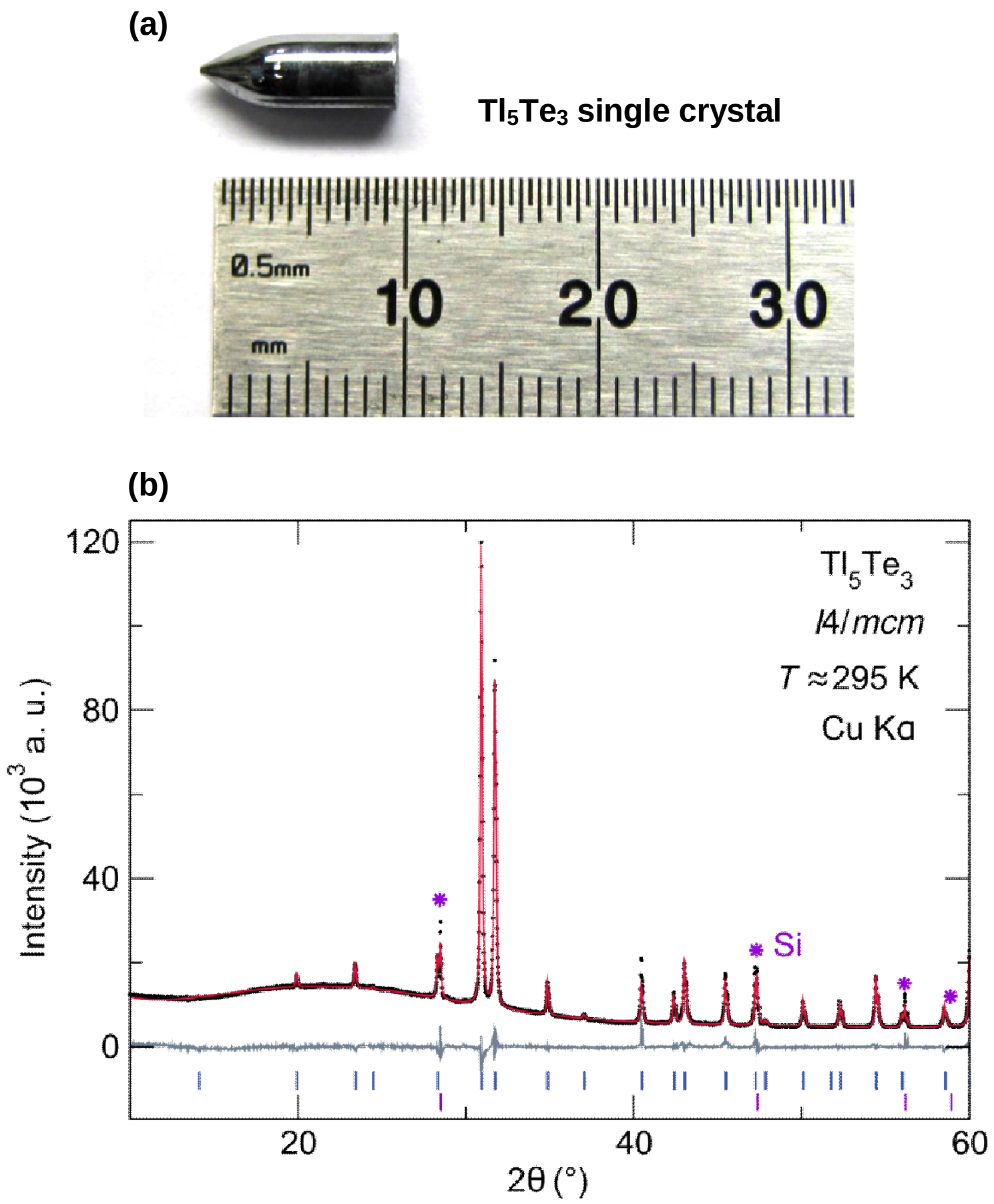

**Fig. 5.** (a) $Tl_5Te_3$ single crystal grown by the optical heating traveling solvent technique using Te as the flux. (b) Powder x-ray diffraction data (black circles) is fit (red line) by Rietveld refinement, with the difference curve in gray. Peaks from a Si standard are marked with purple asterisks. Blue tick marks denote reflections allowed by the *I*4/*mcm* spacegroup of $Tl_5Te_3$, and purple tick marks denote reflections allowed for the Si standard.